\documentclass[Physsubmission, Phys]{SciPost}

\hypersetup{
    colorlinks,
    linkcolor={red!50!black},
    citecolor={blue!50!black},
    urlcolor={blue!80!black}
}

\usepackage[bitstream-charter]{mathdesign}
\DeclareSymbolFont{usualmathcal}{OMS}{cmsy}{m}{n}
\DeclareSymbolFontAlphabet{\mathcal}{usualmathcal}

\newcommand{\beq}{\begin{equation}}

\newcommand{\eeq}{\end{equation}}

\newcommand{\bea}{\begin{eqnarray}}
\newcommand{\eea}{\end{eqnarray}}
\newcommand{\bfig}{\begin{figure}}
\newcommand{\efig}{\end{figure}}
\newcommand{\bc}{\begin{center}}
\newcommand{\ec}{\end{center}}

\begin{document}

\begin{center}{\Large \textbf{
A differential-geometry approach to operator mixing in massless QCD-like theories and Poincar\'e-Dulac theorem\\
}}\end{center}

\begin{center}
Matteo Becchetti
\end{center}

\begin{center}
Physics Department, Torino University and INFN Torino, Via Pietro Giuria 1, I-10125 Torino, Italy
\\
* matteo.becchetti@unito.it
\end{center}

\begin{center}
\today
\end{center}


\definecolor{palegray}{gray}{0.95}
\begin{center}
\colorbox{palegray}{
  \begin{tabular}{rr}
  \begin{minipage}{0.1\textwidth}
    \includegraphics[width=35mm]{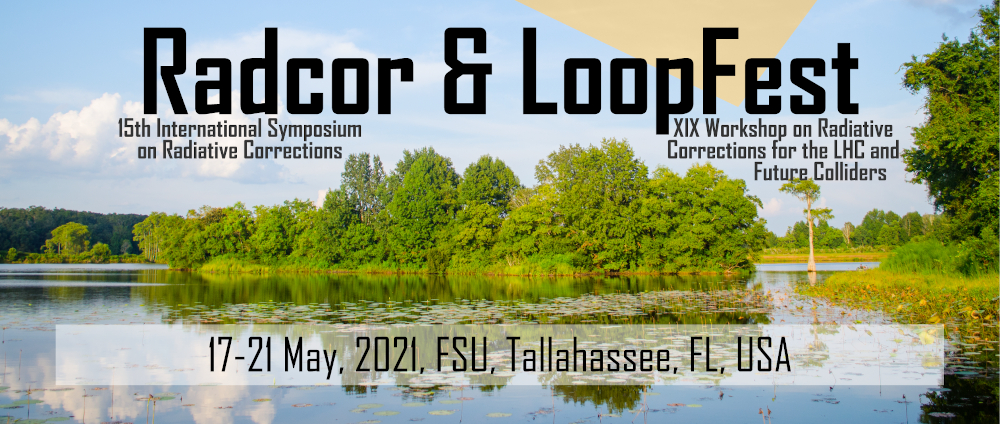}
  \end{minipage}
  &
  \begin{minipage}{0.85\textwidth}
    \begin{center}
    {\it 15th International Symposium on Radiative Corrections: \\Applications of Quantum Field Theory to Phenomenology,}\\
    {\it FSU, Tallahasse, FL, USA, 17-21 May 2021} \\
    \doi{10.21468/SciPostPhysProc.?}\\
    \end{center}
  \end{minipage}
\end{tabular}
}
\end{center}

\section*{Abstract}
{\bf
We review recent progress on operator mixing \cite{MB0,MBP} in the light of the theory of canonical forms for linear systems of differential equations and, in particular, of the Poincar\'e-Dulac theorem. We show that the matrix $A(g) = -\frac{\gamma(g)}{\beta(g)} =\frac{\gamma_0}{\beta_0}\frac{1}{g} + \cdots $ determines which different cases of operator mixing can occur, and we review their classification. We derive a sufficient condition for $A(g)$ to be set in the one-loop exact form $A(g) = \frac{\gamma_0}{\beta_0}\frac{1}{g}$. Finally, we discuss the consequences of the unitarity requirement in massless QCD-like theories, and we demonstrate that $\gamma_0$ is always diagonalizable if the theory is conformal invariant and unitary in its free limit at $g =0$.
}

\vspace{10pt}
\noindent\rule{\textwidth}{1pt}
\tableofcontents\thispagestyle{fancy}
\noindent\rule{\textwidth}{1pt}
\vspace{10pt}

\section{Introduction}
\label{sec:intro}
We reconsider \cite{MB0,MBP} the operator mixing in massless QCD-like theories by exploiting the theory of canonical forms \cite{PD0,PD1} for linear systems of  differential equations, and in particular the Poincar\'e-Dulac theorem \cite{PD1}. We show that operator mixing is characterized  by the structure of the matrix $A(g) = - \frac{\gamma(g)}{\beta(g)}$, where $\gamma(g)$ is the anomalous dimension matrix and $\beta(g)$ is the beta function of the theory.\par
The problem of determining under which conditions the operator mixing reduces to the multiplicatively renormalizable case has been addressed in \cite{MB0} by means of the Poincar\'e-Dulac theorem. This is the case (I) of the classification introduced in \cite{MB0}. The remaining cases, (II),(III) and (IV), of the aforementioned classification, where such a reduction is not actually possible, have been discussed in \cite{MBP}, which also contains physical applications based on \cite{BB1,Buras3,G2,G1}.\par
In the present paper we discuss the operator mixing as it has been worked out in \cite{MB0,MBP} as opposed to previous treatments. \par
It has been known for some time that changing renormalization scheme may provide several advantages. For example, in the 't Hooft renormalization scheme \cite{HH}, all the coefficients of the beta function, $\beta(g) = - \beta_0 g^3 - \beta_1 g^5 + \cdots$, but the first two, $\beta_0, \beta_1$, may be set to $0$ by a suitable holomorphic reparametrization of the gauge coupling. Moreover, the freedom to set to any value the coefficients $\beta_2,\beta_3,\cdots$ has been exploited in various contexts \cite{Kataev1,Kataev2,Kataev3,Kataev4,Kataev5,Kataev6}, including the supersymmetric one in relation to the exact NSVZ beta function \cite{NSVZ}.  As for $\beta(g)$, also for the anomalous dimension of a multiplicatively renormalizable operator, $\gamma(g) = \gamma_0 g^2+\cdots$, it is possible to set to $0$ all the coefficients but the first one, $\gamma_0$,  by a similar \cite{Collins} reparametrization of the coupling. \par
In the mixing case the anomalous dimension is a matrix-valued function. A crucial quantity that enters the operator mixing is the renormalized mixing matrix in the coordinate representation $Z(x,\mu)$. $Z(x,\mu)$ is determined by the quantity $A(g) = - \frac{\gamma(g)}{\beta(g)}$ by means of Eq. \eqref{Zsol}. Consequently, in order to assess whether any simplification of the mixing structure may occur, we exploit \cite{MB0,MBP}  holomorphic gauge transformations of the basis of operators that mix under renormalization, under which $A(g)$ transforms as a meromorphic connection \cite{MB0,MBP}. Therefore, the change of scheme that we refer to in our treatment of operator mixing is deeply different from the aforementioned change of scheme of 't Hooft type. \par
Another aspect that we are going to discuss is a unitarity constraint \cite{MBP}. If we assume that the theory is conformal and unitary at zero coupling in the gauge-invariant sector, as it should be the case for a massless QCD-like theory, then the matrix $\gamma_0$, and therefore also $A_0 = \frac{\gamma_0}{\beta_0}$, should always be diagonalizable and the cases (III) and (IV) of the classification above, where $A_0$ is not diagonalizable, are ruled out.\par
The approach to operator mixing based on the Poincar\'e-Dulac theorem allows us to study in the most general case the UV asymptotics of the renormalized mixing matrix $Z(x,\mu)$ \cite{MB0,MBP}, which enters a number of applications of the renormalization group (RG), ranging from the deep inelastic scattering \cite{Buras1} in QCD to the evalutation of the ratio $\frac{\epsilon'}{\epsilon}$ \cite{Buras2, Buras3, Martinelli} for the possible implications of new physics, and to the constraints \cite{MBM,MBN,MBH,MBR,MBL,BB,BB1} to the eventual nonperturbative solution of the large-$N$ limit  \cite{H,V,Migdal,W} of massless QCD-like theories.\par
The present paper is organized as follows: in section \ref{4} we discuss the classification of operator mixing based on the Poincar\'e-Dulac theorem, in section \ref{5.4} we demonstrate that $\gamma_0$ is diagonalizable in a theory that is conformal and unitary at zero coupling, and finally in section \ref{conc} we state our conclusions.

\section{Operator mixing and the Poincar\`e-Dulac theorem} \label{4}

A differential-geometry interpretation \cite{MB0,MBP} is crucial to exploit in the context of operator mixing the theory of canonical forms for systems of differential equations and the Poincar\'e-Dulac theorem.\par 
Given a system of local operators $O_i(x)$ that mix under renormalization, we interpret \cite{MB0} a change of operator basis, i.e. a change of renormalization scheme:
\bea \label{b}
O'_i(x)=S_{ik}(g) O_k(x)
\eea
as a holomorphic invertible gauge transformation $S(g)$. In this context, the mixing matrix $Z(x,\mu)$ is seen as a Wilson line that transforms as:
\bea
Z'(x, \mu)= S(g(\mu)) Z(x, \mu) S^{-1}(g(x))
\eea
for the gauge transformation $S(g)$. Moreover, $Z(x,\mu)$ is a solution:
\bea \label{Zsol}
Z(x, \mu)=P\exp\left(\int ^{g(\mu)}_{g(x)} A(g) \, dg\right)
=P\exp\left(-\int^{g(\mu)}_{g(x)}\frac{\gamma(g)}{\beta(g)}dg\right)
\eea
of the differential system:
\bea \label{1.700}
\mathcal{D} Z(x,\mu)=  \left(\frac{\partial}{\partial g} - A(g)\right) Z(x,\mu) =0
\eea
From Eq. \eqref{Zsol} we see that the structure of $Z(x,\mu)$, and therefore the structure of the operator mixing, is determined by the matrix:
\bea \label{A}
A(g)=-\frac{\gamma(g)}{\beta(g)}= \frac{\gamma_0}{\beta_0} \frac{1}{g}+\cdots
\eea
which is interpreted \cite{MB0} as a meromorphic connection, with a simple pole at $g=0$,
that transforms by the gauge transformation $S(g)$ as:
\bea
A'(g)= S(g)A(g)S^{-1}(g)+ \frac{\partial S(g)}{\partial g} S^{-1}(g)
\eea

\subsection{The Poincar\`e-Dulac theorem for massless QCD-like theories} 

Once the above geometrical framework for operator mixing has been established, we are able to discuss our application of the Poincar\'e-Dulac theorem. \par
Naively, given a linear system of differential equations of the kind \eqref{1.700}, the Poincar\'e-Dulac theorem establishes which type of simplifications can be made on the meromorphic connection $A(g)$ by means of holomorphic gauge transformations.

  In a massless QCD-like theory the connection $A(g)$ admits the expansion \cite{MB0}:
\bea \label{sys}
A(g)=-\frac{\gamma(g)}{\beta(g)}= \frac{1}{g} \left(A_0 + \sum^{\infty}_ {k=1} A_{2k} g^{2k} \right)
\eea
According to the Poincar\`e-Dulac theorem \cite{PD1}, $A(g)$ can be set by a holomorphic invertible gauge transformation in the canonical resonant form \cite{MBP}:
\bea \label{resft}
A'(g)=\frac{1}{g} \left(\Lambda+N_0 +  \sum_{k=1} N_{2k} g^{2k} \right)
\eea
where:
\bea \label{J}
A_0=\Lambda+N_0
\eea
is upper triangular, with eigenvalues $\text{diag}(\lambda_1, \lambda_2, \cdots)=\Lambda$ in nonincreasing order $\lambda_1 \geq \lambda_2 \geq \cdots$, and nilpotent part, $N_0$, in normal Jordan form. The upper triangular nilpotent matrices $N_{2k}$ satisfy the condition:
\bea \label{nil}
g^{\Lambda} N_{2k} g^{-\Lambda} = g^{2k} N_{2k}
\eea
This implies \cite{MBP} that the only nonzero entries $(N_{2k})_{ij}$ are such that:
\bea \label{ein1}
\lambda_i - \lambda_j=2k
\eea
for $i < j$ and $k$ a positive integer. Eq. \eqref{ein1} is called the \emph{resonance condition} for the eigenvalues of $A_0$, and we refer to the terms $N_{2k}$ in Eq. \eqref{resft} as to the \emph{resonant terms}. A similar condition also is satisfied by the nilpotent matrix $N_0$:
\bea
g^{\Lambda} N_{0} g^{-\Lambda} =  N_{0}
\eea
as a consequence of the Jordan normal form of $A_0$. \par
Therefore, the Poincar\'e-Dulac theorem establishes under which conditions the connection $A(g)$ may be set in the form:
\bea \label{eul}
A'(g)= \frac{\Lambda+N_0}{g} 
\eea
by a holomorphic gauge transformation. In particular, a sufficient condition for all the resonant terms $N_{2k}$ to be absent in Eq. \eqref{resft}
is that the eigenvalues of $A_0$ satisfy \cite{MB0}:
\bea \label{ein}
\lambda_i - \lambda_j \neq 2k
\eea
with $k$ a positive integer. Remarkably, this condition can be easily verified a priori from the only knowledge of the eigenvalues of
$A_0 = \frac{\gamma_0}{\beta_0}$, i.e. it is possible to test the condition \eqref{ein1} performing a one-loop computation only:
\begin{center}
\emph{If the condition \eqref{ein} is satisfied for all the eigenvalues of $A_0$, then $A(g) = - \frac{\gamma(g)}{\beta(g)}$ may be set in the one-loop exact form \eqref{eul} to all orders in perturbation theory for a choice of the operator basis}.
\end{center} 
Moreover, this analysis shows the crucial difference with respect to the multiplicatively renormalizable case: While in the multiplicatively renormalizable case it is always possible to remove from the perturbative expansion of $A(g)$ all the coefficients different from $\frac{\gamma_0}{\beta_0}$, in the mixing case this is only possible if either the condition \eqref{ein} is satisfied for all the eigenvalues of $A_0$ or $(N_{2k}) = 0$ for all $k \geq 1$.

\subsection{Classification of operator mixing}

We summarize the classification of operator mixing in \cite{MB0,MBP}:
\begin{itemize}
\item \textbf{(I) Nonresonant diagonalizable $\frac{\gamma_0}{\beta_0}$}:\par
The system in Eq. \eqref{1.700} of differential equations associated to the connection $A(g)$ is nonresonant and $\frac{\gamma_0}{\beta_0}$ is diagonalizable.
For the system to be nonresonant, it is sufficient that the eigenvalues of $\frac{\gamma_0}{\beta_0}$ satisfy:
\bea
\lambda_i - \lambda_j \neq 2k
\eea
with  $i \leq j$ and $k$ a positive integer.\par
For the system to be nonresonant, the necessary and sufficient condition is that in the canonical form of Eq. \eqref{resft} all the resonant terms vanish.\par
The sufficient condition for $\frac{\gamma_0}{\beta_0}$ to be diagonalizable is that all its eigenvalues are different.
\item  \textbf{(II) Resonant diagonalizable $\frac{\gamma_0}{\beta_0}$}:\par
The system of differential equations is resonant and $\frac{\gamma_0}{\beta_0}$ is diagonalizable. For the system to be resonant, a necessary condition is that for at least two eigenvalues it holds:
\bea 
\lambda_i - \lambda_j = 2k
\eea
with $i < j$ and $k$ a positive integer. \par
In this case, a necessary and sufficient condition is that at least one $N_{2k}$ in the canonical resonant form does not vanish.\par
The sufficient condition for $\frac{\gamma_0}{\beta_0}$ to be diagonalizable is as in the case (I).
\item  \textbf{(III) Nonresonant nondiagonalizable $\frac{\gamma_0}{\beta_0}$}:\par
The system of differential equations is nonresonant and $\frac{\gamma_0}{\beta_0}$ is nondiagonalizable.\par 
The nonresonant condition is as in the case (I). \par
The necessary condition for $\frac{\gamma_0}{\beta_0}$ to be nondiagonalizable is that at least two of its eigenvalues coincide.
\item  \textbf{(IV) Resonant nondiagonalizable $\frac{\gamma_0}{\beta_0}$}:\par
The system of differential equations is resonant and $\frac{\gamma_0}{\beta_0}$ is nondiagonalizable.\par
The resonant condition is as in the case (II).\par
The necessary condition for $\frac{\gamma_0}{\beta_0}$ to be nondiagonalizable is as in the case (III).
\end{itemize}

\section{Unitarity constraint} \label{5.4}

We discuss the unitarity constraint \cite{MBP} that rules out the cases (III) and (IV) of the classification above.  \par
We observe that conformal invariance and unitarity should apply to massless QCD-like theories up to the order of $g^2$ in perturbation theory. Indeed, since the beta function affects the solution of the Callan-Symanzik equation \cite{C,S} starting from the order of $g^4$, massless QCD-like theories are conformal invariant up to the order $g^2$ \cite{ConfQCD}. Moreover, unitarity is certainly satisfied in the free limit of massless QCD-like theories in the Hermitian gauge-invariant sector, as unitary gauges exist where the theory is unitary for the gluon and matter fields and the gauge-fixing ghosts decouple in the correlators of gauge-invariant operators.\par
We demonstrate in the following, exploiting unitarity and conformal invariance, that the matrix $\gamma_0$ is always diagonalizable in the gauge-invariant sector of massless QCD-like theories and therefore the cases (III) and (IV) of our classification are ruled out.\par
To prove the previous statement, we work in the framework of conformal field theories (CFTs) and we study the Callan-Symanzik equation satisfied by the $2$-point correlators of Euclidean Hermitian scalar primary conformal operators $G_{conf}(x)$ \cite{MBP}:
\beq
\label{CSconf}
x\cdot\dfrac{\partial}{\partial x}G_{conf}(x) + \Delta G_{conf}(x) + G_{conf}(x)\Delta^T = 0
\eeq
where $\Delta$ is the matrix of the conformal dimensions.\par
The general solution of Eq. \eqref{CSconf}, in matrix notation, is \cite{MBP}:
\beq
\label{G2conf}
G_{conf}(x) = \langle O(x)O(0) \rangle = e^{-\Delta \log \sqrt {x^2 \mu^2}} \mathcal{G} e^{-\Delta^T \log \sqrt {x^2 \mu^2}}
\eeq
 where $\mathcal{G}$ is a real symmetric matrix. \par
The scalar product of the theory is constructed by exploiting the operators/states correspondence in CFTs \cite{Hog,CFT2}: 
\begin{eqnarray}
O(0)\vert 0 \rangle &=& \vert O_{in}\rangle \nonumber \\
\langle O_{out}\vert &=& \lim_{x\rightarrow \infty}\langle 0\vert e^{2\Delta\log \sqrt {x^2 \mu^2}} O(x)
\end{eqnarray}
As a consequence, the scalar product in matrix notation is:
\begin{eqnarray}
\label{ScalarP}
\langle O_{out}\vert O_{in} \rangle &=& \lim_{x\rightarrow \infty} \langle 0 \vert e^{2\Delta\log \sqrt {x^2 \mu^2}} O(x)O(0)\vert 0 \rangle \nonumber \\
& = &\lim_{x\rightarrow \infty}  e^{2\Delta\log \sqrt {x^2 \mu^2}} e^{-\Delta \log \sqrt {x^2 \mu^2}} \mathcal{G} e^{-\Delta^T \log \sqrt {x^2 \mu^2}}  \nonumber \\
&=& \lim_{x\rightarrow \infty}  e^{\Delta\log \sqrt {x^2 \mu^2}} \mathcal{G} e^{-\Delta^T\log \sqrt {x^2 \mu^2}} 
\end{eqnarray}
In order to be well defined, the scalar product in Eq. \eqref{ScalarP} must be independent of the space-time coordinates.
Expanding the exponentials in the last line of Eq. \eqref{ScalarP}, we get:
\begin{eqnarray}
\label{2.34}
\langle O_{out} \vert O_{in} \rangle &&= \left(I +\Delta  \mathcal{G}\log \sqrt {x^2 \mu^2} +\cdots\right)\mathcal{G}\left(I -   \mathcal{G} \Delta^T \log \sqrt {x^2 \mu^2} +\cdots\right) \nonumber \\
&&=   \mathcal{G} +\left(\Delta  \mathcal{G} -  \mathcal{G} \Delta^T \right) \log \sqrt {x^2 \mu^2} +\cdots
\end{eqnarray}
Then, the independence of the coordinates implies the relation \cite{MBP}:
\bea \label{main}
\Delta  \mathcal{G} -  \mathcal{G} \Delta^T=0
\eea
We apply the previous analysis to massless QCD-like theories, since they are conformal invariant in perturbation theory up to the order of $g^2$ and, in particular, the following perturbative expansions hold:
\begin{eqnarray} \label{pert11}
\Delta(g) & = &  D \, I + g^2 \gamma_0 + \cdots \\ \nonumber
\mathcal{G}(g) & = & G_0 + g^2 G_1 + \cdots
\end{eqnarray}
in the conformal renormalization scheme \cite{ConfQCD}, where $D$ is the canonical dimension of the operators $O_i$ that mix under renormalization. Therefore, expanding Eq. \eqref{main} to the order of $g^2$, we obtain \cite{MBP}:
\begin{eqnarray}
\label{2.35}
\gamma_0 G_0 - G_0\gamma_0^T=0
\end{eqnarray}
We distinguish two cases \cite{MBP}:
\begin{itemize}
\item \textbf{(I) $\gamma_0$ is diagonalizable}: Eq. \eqref{2.35} implies that $G_0$ commutes with $\gamma_0$ in the diagonal basis and thus in any basis. Moreover, since $G_0$ is a real symmetric matrix, it is diagonalizable as well, and therefore $G_0$ and $\gamma_0$ are simultaneously diagonalizable. Moreover, if the theory is unitary, $G_0$ must have positive eigenvalues.
\item \textbf{(II) $\gamma_0$ is nondiagonalizable}:  In this case $G_0$ has necessarily negative eigenvalues, hence the theory is nonunitary in its free conformal limit at $g=0$ in the gauge-invariant Hermitian sector.  \par
Indeed, if $\gamma_0$ is nondiagonalizable, Eq. \eqref{main} nontrivially constrains \cite{Hog}\cite{MBP} the structure of $G_0$, which can be set in the canonical form  \cite{Hog}\cite{MBP}:
\beq \label{GantiDiag}
G'_0 = \left(\begin{array}{ccccc}
0 & 0 & 0 & \cdots & 1 \\
0 & 0 & \iddots & 1 & 0 \\
0 & \iddots & 1 & 0 & 0 \\
\vdots & \iddots & \vdots & \vdots & \vdots \\
1 & 0 & \cdots & \cdots
 & 0
\end{array}\right)
\eeq
by a suitable change of basis. It turns out \cite{Hog}\cite{MBP} that $G'_0$ has $[r/2]$ positive eigenvalues and $[r/2]$ negative eigenvalues, where $r$ is the rank of $G_0'$. Therefore, the scalar product of the theory, which is induced by $G'_0$, is not positive definite and, consequently, the theory is nonunitary.
\end{itemize}
By summarizing, the nondiagonalizability of $\gamma_0$ and the existence of the conformal structure to the order of $g^2$
determine the structure of $G_0$, which defines the scalar product in the free conformal limit, in such a way that the free conformal limit is nonunitary if $\gamma_0$ is nondiagonalizable. Therefore, the perturbative conformal symmetry to the order of $g^2$, and the lowest-order unitarity, rule out the cases (III) and (IV) of operator mixing in the gauge-invariant Hermitian sector of a massless QCD-like theory. However, the previous statement does not necessarily apply outside the gauge-invariant sector that need not to be unitary due to the mixing with the ghost sector.

\section{Conclusion}  \label{conc}

In the present paper we have reviewed recent progress on the problem of the operator mixing in massless QCD-like theories \cite{MB0,MBP} that exploits the theory of canonical forms for linear systems of differential equations and, specifically, the Poincar\'e-Dulac theorem. We have shown that the differential-geometry perspective \cite{MB0,MBP} allows us to obtain a complete classification of operator mixing, and to assess whether or not it may be reduced to the multiplicatively renormalizable case by a suitable choice of the operator basis. Finally, imposing unitarity of the free conformal limit, we have demonstrated that the first coefficient, $\gamma_0$, of the anomalous dimension matrix should always be diagonalizable in the gauge-invariant sector of a massless QCD-like theory.

\section*{Acknowledgements}

M.B. acknowledges the financial support from the European Union Horizon 2020 research and innovation programme: \emph{High precision multi-jet dynamics at the LHC} (grant agreement no. 772009).




\nolinenumbers

\end{document}